\documentstyle[12pt]{article}

\author
{S. N. Dorogovtsev \thanks{e-mail: sn@dor.ioffe.rssi.ru}\\
{\it A. F. Ioffe Physico-Technical Institute,} \\
{\it 194021, St. Petersburg, Russia}}
\title{Gyrotropic Percolation
}

\date{}

\begin{document}

\maketitle

\renewcommand{\abstractname}{}
\begin{abstract}
%\noindent
Spiral (gyrotropic) percolation which is related to the behavior
of an electron system in strong magnetic fields is studied. It is
shown that the scaling behavior area near the percolation threshold
is anomalously narrow. The percolation threshold is higher than
in a system with usual isotropic percolation (i.e., at higher
concentrations of undamaged structure elements). Our old value of the
critical exponent of the correlation length is corrected.
\end{abstract}

The following is the translation from Russian of my old paper (1986)
published in a poorly known (though it is translated into
English \cite{a1}) Russian journal \cite{a1}.
For this reason the paper seems
to escape notice. (Note that Eq. (\ref{e1}) is written in a
more compact form and misprints are corrected, a style of the translation
is also corrected.)

Before the beginning of the main text, I want to mention several very
important subsequent papers on the spiral percolation
\cite{b1} (P. Ray and I.~Bose) and \cite{b2,b3,b4}
(S. B. Santra and I.~Bose).
These authors have come to idea of this type of percolation from their
studies of spiral lattice animals \cite{b5,b6,b7,b8}.
One can find modern results -- the best values of critical exponents and
the thresholds for different lattices obtained by direct calculations --
and fresh references concerning the problem
under consideration in the mentioned papers and in interesting lectures
of I.~Bose \cite{b9}. I am
grateful to Prof. I.~ Bose who have informed me about these valuable papers.
Note that all other references are only to papers published before
\cite{a1}.

The probability of electron jumps in one or another direction with respect
to the direction of the preceding jump is anisotropic function of the
corresponding angle \cite{r1,r2} for systems with hopping conduction in
an applied magnetic field. This anisotropy is enhanced in classically
strong fields and can lead to changes in the percolation properties of the
material. We shall consider the simplest lattice model to study such
systems.

Let us consider, for example, the bond problem for a two-dimensional
lattice. A particle is allowed to move from site to site via bonds and a
fraction of such bonds is broken randomly. We shall assume that each
subsequent displacement of the particle via allowed bonds relative to its
preceding displacement can take place only in two ways: forward or to the
left, i.e., right-hand turns are forbidden.

If the direction of each
subsequent step were determined by a random-number generator, we would have
a problem of a spiral random walk (see, for example, \cite{r3}). As for an
ordinary random walk \cite{r4,r5}, we can study a random walk of this type
also in a directed system. However, we shall consider a different problem,
i.e., what values of the concentration of the unbroken bonds are required
for a particle to pass from one edge of a lattice to the other one if we
forbid the right-hand turns. We are also interested in the critical
properties of such a process which is called gyrotropic (or spiral)
percolation. (Of course, systems with higher dimensionality can also be
considered.)

The crudest naive estimate of the percolation threshold for gyrotropic
percolation on a hypercubic lattice of dimensionality $D$ yields
$u_c \sim u_{c0}[1-1/(2D-1)]^{-1}$, where $u$ is the concentration of the
unbroken bonds and $u_{c0}$ is the percolation threshold for ordinary
isotropic percolation. Enumerating random configurations on a square
lattice (the bond problem), we obtain $u_c = 0.69 \pm 0.03$
$(u_{c0}=0.5)$.

We shall use the real space renormalization group approach in our further
consideration (position space renormalization group). The
percolation
problem for a system of conductors and randomly oriented diodes (randomly
directed percolation) was first studied by S. Redner \cite{r6,r7,r8,r9}.
Such a problem is a generalization of the directional percolation problem
\cite{r10,r11,r12}.
It will be shown here, that in the case under consideration, the
renormalization group transformations induce randomly directed bonds like
bonds ab initio presenting on a lattice for the randomly directed
percolation problem.

The renormalization group transformations transform clusters consisting
of several bonds to new effective bonds, so in fact we apply sequentially
a coarsening procedure. It follows from Fig.~1 that even a small cluster
on a square lattice contains configurations of isotropic bonds which
induce upon renormalization transformation oriented bonds.
Let us denote by $q$ the probability of appearance of such bonds. The
probability of finding a broken bond is then given by $1-u-2q$, since
there are possible two bond orientations. Of course, the bare value of the
above probability should be $q(0)=0$!

To make the problem completely local, we assume that {\it a particle can not
come to a stop and begin to move backwards}. Note, it is a principal
assumption of the problem under consideration!

To obtain the required recursive renormalization group relations for the
probabilities defined above, we have to enumerate $(1+1+2)^N$ possible
configurations, where $N$ is the number of bonds in the cluster used.
Large clusters containing at least several tens of bonds are required for
quantitative analysis. Thus, the enumeration is possible only by the Monte
Carlo method and a large computer is required.

Since we are interested only
in qualitative features of this type of percolation, it is sufficiently
to consider a small cluster consisting of six bonds on a hierarchical
lattice (see Fig.~2). Even using this minimal
suitable cluster (available for manual enumeration of configurations), one
can find that the renormalization transformations induce oriented bonds.
Although the cluster considered does not give the exact isotropic percolation
threshold (like a cluster consisting of four bonds with a scaling
transformation coefficient $1/\sqrt2$), but enables us to obtain after the
recursive transformation both oriented and nonoriented bonds.

Such a cluster admits a large enough number of percolation channels. This
requirement is not satisfied by the simplest self-dual cluster consisting
of five bonds \cite{r11}. The main features of the corresponding recursive
relations and the qualitative positions of their fixed points remain
unchanged for larger clusters.

Let us write the recursive relations immediately (in \cite{a1} these
relations were presented in an expanded form):
\begin{eqnarray}
\label{e1}
u^\prime=u^2 [1-(u+q)^2]^2 + (u+q)^4 [3-2(u+q)^2]
\nonumber      \ , \\[2ex]
q^\prime=[1-(u+q)^2]^2 [2(u+q)^2 - u^2]  \ . \ \ \ \ \ \ \ \ \ \ \ \ \ \ \ \
\end{eqnarray}

The corresponding phase diagram is shown in Fig.~3. The gyrotropic
percolation threshold is $u_c=(\sqrt5-1)/2=0.618$. Using the same cluster
for isotropic percolation, one obtains the following recursive relation:
\begin{equation}
\label{e2}
u^\prime=1-(1-u^2)^3  \ ,
\end{equation}

\noindent
and $u_{c0}=(3-\sqrt5)/2=\sqrt{u_c}=0.389$. Such a shift of the
percolation threshold leads to lowering of a conductivity of the structure
relative to the usual one. This reduction corresponds to increasing of
the magnetoresistance with increasing magnetic field \cite{r13}.

The recursive
relations (\ref{e1}) have only three fixed points: $(0,0),\ (1,0)$, and
$(u^*,q^*)=((3-\sqrt5)/2,\sqrt5-2)=(0.382,0.236)$. The equation for one of
the separatrices is $u+q=(\sqrt5-1)/2=0.618$. The coordinates of the
fixed point $(u^*,q^*)$ are also related by $2u^*+q^*=1$. The maximum of
the second separatrix $q=\sqrt{u}-u$ is at the point
$(u_{max},q_{max})=(1/2,1/4)$.

Beautiful phase diagrams very similar to that shown in Fig.~3
but much more symmetrical ones were first
obtained in papers of S. Redner and coauthors where randomly directed
percolation was studied. However, in the Redner's problem, oriented bonds
are not induced by the renormalization if bare $q(0)$ is zero.
In contrast to problem \cite{r6,r7,r8,r9} the threshold point under
consideration $(u_c,0)$ is not a fixed point. In fact, the critical
behavior is determined by the fixed point $(u^*,q^*)$. It is easy to see
that one gets after n transformations (\ref{e1}) the following deviation
\begin{equation}
\label{e3}
u(n)-u^*=C(n)(u(n=0)-u_c)  \ .
\end{equation}

\noindent
Here, $C(n)$ depends only on $n$, $u(n=0)-u_c \ll 1$, and $n$ is chosen
such one that $u(n)-u^* \ll 1$. Thus, the scaling exponent $\nu$
for the correlation length $\xi \sim |u-u_c|^{-\nu}$ can be found by
linearization of Eq. (\ref{e2}) near the fixed point $(u^*,q^*)$:
\begin{eqnarray}
\label{e4}
\left(
\begin{array}{c}
u^\prime - u^* \\
q^\prime - q^*
\end{array}\right) =
\left(
\begin{array}{cc}
13-5\sqrt5   & 2(3-\sqrt5) \\
-(7-3\sqrt5) & 0
\end{array}
\right)
\left(
\begin{array}{c}
u^\prime - u   \\
q^\prime - q
\end{array}\right)  \nonumber  \\[2ex]
=
\left(
\begin{array}{cc}
1.820  & 1.528 \\
-0.292 & 0
\end{array}
\right)
\left(
\begin{array}{c}
u^\prime - u   \\
q^\prime - q
\end{array}\right)  \ .  \ \ \ \ \ \ \ \ \ \
\end{eqnarray}

\noindent
and $\nu \propto 1/\log(\lambda_1)$. Here $\lambda_1=2(3-\sqrt5)=1.528$
and $\lambda_2=7-3\sqrt5=0.292$ are exigent values of Eq. (\ref{e4}).
The value $\lambda=\partial u^\prime/\partial u (u_{c0})=1.681$ may be
obtained for isotropic percolation using the {\it same} cluster
(see Eq. (\ref{e2})). Thus The critical exponent $\nu$ for gyrotropic
percolation is 1.225 of $\nu$ value for ordinary percolation.

For the unrenormalized values of $u(0)$ satisfying $u(0)-u_c > 10^{-2}$,
the phase trajectory passes far from the fixed point $(u^*,q^*)$ and
Eq. (\ref{e4}) will never be valid. Thus, the scaling region is much
narrower than for isotropic percolation. Such an anomalously narrow region
hardly can be observed for realistic systems. A similar narrowing of the
scaling region should occur for systems in which the renormalization
transformation induce bonds of new type.

We have studied only a strongly gyrotropic situation. Under more realistic
conditions  when right-hand turns are possible although less probable than
left-hand turns, the increase of the percolation threshold is not so large
but should be noticeable.

Finally, we would like to point out that larger clusters cutted from
realistic lattice systems could lead to different shapes of separatrices
as in the case of randomly directed percolation \cite{r6,r7,r8,r9}.

\medskip
The author thanks V. V. Bryksin, A. V. Goltsev,
E.~K.~Kudinov, and V.~N.~Prigodin for many helpful discussions.
The author grateful especially to S.~Redner who pointed out him to the
principal papers on randomly oriented diode networks. The author grateful
to Prof. I.~Bose for kind information about very important papers on the
spiral percolation \cite{b1,b2,b3,b4} and spiral
animals \cite{b5,b6,b7,b8}.

%\newpage
%\renewcommand{\refname}{}

\newpage

\begin{center}
FIGURE CAPTIONS \\
\end{center}
\vspace{8mm}

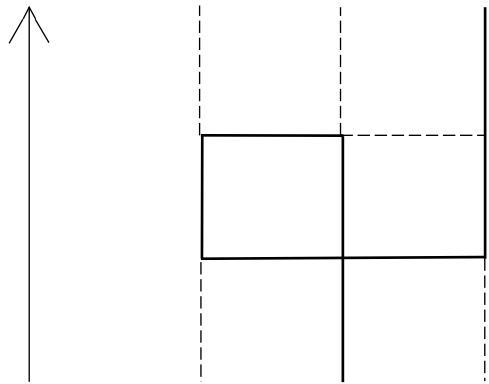
Fig. 1.  Example of a configuration which induces oriented renormalized
bonds in the problem of gyrotropic percolation. Right-hand turns are
forbidden.  The broken lines denote broken bonds in a cluster. The arrow
indicates possible orientation of percolation.

\vspace{8mm}

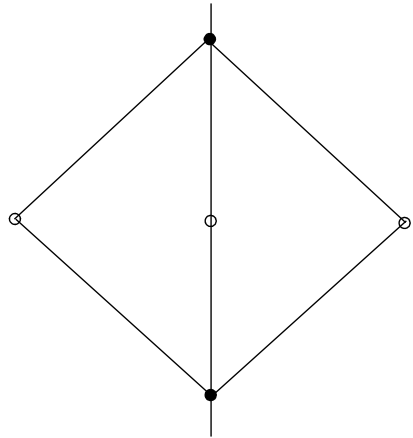
Fig. 2.  Clusters used at the derivation of recursive relations (1) and
(2).

\vspace{8mm}

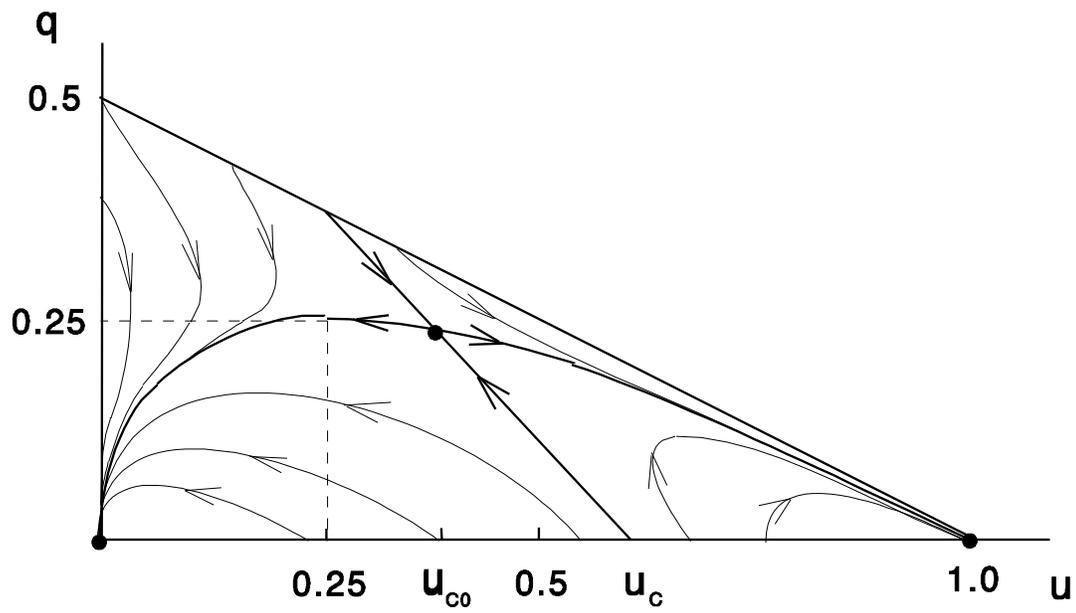
Fig. 3. Phase diagram of the recursive relations (1). Compare with the
corresponding phase diagrams of S. Redner et al
for randomly directed percolation \cite{r6,r7,r8,r9}.

\end{document}